\documentclass[preprintnumbers,floatfix,preprint,prc]{revtex4}

\usepackage{enumerate}
\usepackage{amsmath,amssymb}  
\usepackage{bm}               
\usepackage{amscd}            
\usepackage{graphicx}         
\usepackage{epsfig}
\usepackage{hhline,multirow}  
\usepackage{dcolumn}          
\usepackage{color}
\usepackage[normalem]{ulem}

\newcommand{\nbar}{{\overline n}}
\newcommand{\nslash}{n\hspace*{-0.22cm}\slash\hspace*{0.022cm}}
\newcommand{\nbslash}{\nbar\hspace*{-0.22cm}\slash\hspace*{0.022cm}}
\newcommand{\Tr}{{\rm Tr}}

\begin{document}
\preprint{JLAB-THY-15-2002}

\title{\mbox{Hadron mass corrections in semi-inclusive
		deep-inelastic scattering}}

\author{J.~V.~Guerrero$^{\, 1,2}$,
	J.~J.~Ethier$^{\, 2,3}$,
	A.~Accardi$^{\, 1,2}$,
	S.~W.~Casper$^{\, 2,4}$\footnote{Current address:
		University of Wisconsin, Madison, Wisconsin 53706, USA}
	W.~Melnitchouk$^{\, 2}$}

\affiliation{
$^1$\mbox{Hampton University, Hampton, Virginia 23668, USA} \\
$^2$\mbox{Jefferson Lab, Newport News, Virginia 23606, USA} \\
$^3$\mbox{College of William and Mary, Williamsburg, Virginia 23185, USA} \\
$^4$\mbox{Carnegie Mellon University, Pittsburgh, Pennsylvania 15213, USA} \\
}

\date{\today}

\begin{abstract}
The spin-dependent cross sections for semi-inclusive lepton-nucleon
scattering are derived in the framework of collinear factorization,
including the effects of masses of the target and produced hadron
at finite momentum transfer squared $Q^2$.
At leading order the cross sections factorize into products of
parton distribution and fragmentation functions evaluated in terms
of new, mass-dependent scaling variables.
The size of the hadron mass corrections is estimated at kinematics
relevant for future semi-inclusive deep-inelastic scattering
experiments.
\end{abstract}

\maketitle

\section{Introduction}

Determining the detailed flavor and spin structure of the nucleon
remains a central challenge for hadronic physics into the 21st century.
Considerable progress has been made over the past two decades in
understanding the characteristics of the momentum and spin
distributions of quarks and gluons (or partons) through precise
measurements of the nucleon's parton distribution functions (PDFs)
in various hard scattering reactions \cite{JMO13, FW13}.
In addition to the traditional inclusive deep-inelastic scattering
(DIS), Drell-Yan and other high-energy scattering processes,
an increasingly important role in this quest has been played by
semi-inclusive production of hadrons in lepton--nucleon scattering.

Identification of specific hadrons, such as pions or kaons, in the
current fragmentation region of a deep-inelastic collision serves
as a tag of individual quark flavors, which in inclusive DIS are
summed over.  Important insights have been provided through
semi-inclusive deep-inelastic scattering (SIDIS) experiments on
phenomena such as the SU(2) flavor asymmetry in the proton sea
\cite{HERMES_dubar} and the ratio of strange to nonstrange quark
distributions \cite{HERMES_s}.  From experiments with polarized
targets, SIDIS data have also provided fascinating glimpses of the
possible flavor asymmetry in the polarized light-antiquark sea
\cite{HERMES_polsea}, while kaon production data has fueled the
recent controversy concerning the sign of the polarized strange
sea \cite{HERMES_Deltas, LSS11}.  Furthermore, detection
of forward baryons (in the center of mass frame) in the target
fragmentation region of SIDIS is a potentially important avenue
for extracting information on the pion cloud of the nucleon or
the structure of the virtual pion itself
\cite{Thomas83, MT95, HERA99, TDIS14}.

In more recent developments, detection of both the longitudinal
and transverse momentum distributions of hadrons produced in
SIDIS measurements of various single- and double-spin asymmetries
has opened up the largely unexplored realm of transverse momentum
dependent parton distributions \cite{Mulders96, Bacchetta07, Lefky14}.
These reveal an even richer landscape of three-dimensional momentum
and spin distribution of partons in the nucleon, that will be the
subject of increasingly greater attention at facilities such as
Jefferson Lab \cite{JLab11} and COMPASS \cite{COMPASS14}, and a
central component of the science program at the proposed
Electron-Ion Collider \cite{EIC}.

The unambiguous interpretation of any SIDIS experiment in terms
of leading twist PDFs or transverse momentum distributions requires
control of various subleading $1/Q^2$ corrections, such as target
mass and higher twist effects, as well as knowledge of the
fragmentation functions describing the parton hadronization.
For inclusive DIS, the finite-$Q^2$ corrections are known to become
important at low $Q^2$ values, particularly when the parton momentum
fraction $x$ is large \cite{GP76, Schienbein08, Accardi13}.
Their effects on global fits of spin-averaged PDFs have been
systematically studied in recent analyses by the CTEQ-Jefferson Lab (CJ)
\cite{CJ10, CJ11, CJ12}, ABM \cite{ABKM10, ABM12} and JR groups
\cite{JR14}, and in spin-dependent PDF analyses by the JAM collaboration
\cite{JAM} (and to some extent also by the LSS \cite{LSS10},
BB \cite{BB10} and NNPDF \cite{NNPDF13} groups).

Typically, the effects of target mass corrections (TMCs) can be
computed within a specific framework, while higher twist effects,
which involve more complicated multi-parton correlations, are
parametrized phenomenologically.  The standard approach for computing
TMCs has traditionally been within the operator product expansion,
in which the mass corrections to inclusive DIS structure functions
arise from twist-two quark bilinear operators with an arbitrary
number of derivative insertions \cite{GP76, Nachtmann73}.
Extending this framework to processes involving particles in the
final state is problematic, however, which has in practice limited
the study of hadron mass corrections in SIDIS.

An alternative framework for TMCs was developed using techniques
based on collinear factorization (CF) \cite{EFP, CSS88, CS82}, in
which the hard scattering is formulated in momentum space directly.
The method has been applied to the computation of TMCs in inclusive
scattering, both in unpolarized \cite{KretzerCF, AOT94, AQ08} and
polarized \cite{AM08} DIS, and in semi-inclusive hadron production 
in electron-proton anhilation \cite{AAR15, Albino08}. 
For semi-inclusive hadron production in lepton-proton collisions,
in contrast to inclusive DIS and $e^+ e^-$ anihilation, finite-$Q^2$ 
corrections can arise from both the effects of the target mass 
and the mass of the produced hadron.  While earlier analyses
\cite{Mulders01, Albino07} considered some of these corrections
within the CF framework, the phenomenology of the combined effects of
the target and produced hadron masses -- which we refer collectively
as ``hadron mass corrections'' (HMCs) -- was systematically explored
in Ref.~\cite{AHM09} for unpolarized scattering.

In this work we extend the analysis of HMCs to the case of
spin-dependent SIDIS at finite~$Q^2$.  Because high energy
spin-dependent data are generally more scarce than spin-averaged
cross sections, a significantly larger fraction of the world's
data set used to constrain spin-dependent PDFs lies in the
low-$Q^2$ region ($Q^2 \sim 1-2$~GeV$^2$).
While target mass corrections have been incorporated in some
global spin-PDF analyses \cite{JAM, LSS10, BB10, NNPDF13},
none of the analyses which have included polarized SIDIS data
\cite{LSS10, DSSV09} have accounted for HMCs.
With the increasing precision of new polarized measurements,
and the consequently more accurate determination of spin-dependent
PDFs, it is imperative to reliably account for subleading corrections
which could impact the extraction of the leading twist distributions.

In Sec.~\ref{sec:semi-intro} we outline the formalism used to
compute the SIDIS cross sections at finite values of $Q^2$
within the collinear approximation in the presence of target
and produced hadron masses.  
For completeness, we consider both polarized and unpolarized
scattering, since the latter enters the calculation of the
measured polarization asymmetries.
In Sec.~\ref{ssec:cf} we review the collinear formalism and
its application to hadron production in SIDIS.
Expanding the hadronic tensor in terms of quark correlators,
in Sec.~\ref{ssec:cross} we derive semi-inclusive cross sections,
which at leading order are given by factorized products of
PDFs and fragmentation functions expressed as functions of
modified scaling variables.
The relative importance of the HMCs is explored numerically
in Sec.~\ref{sec:results}, where we quantify the dependence
of the finite-$Q^2$ cross sections on the kinematical variables
and estimate the corrections for specific current and future
experiments.  Finally, we conclude by summarizing our results
in Sec.~\ref{sec:conclusion}.

\section{Semi-inclusive scattering with mass corrections}
\label{sec:semi-intro}

\begin{figure}[t]
\includegraphics[height=4.5cm]{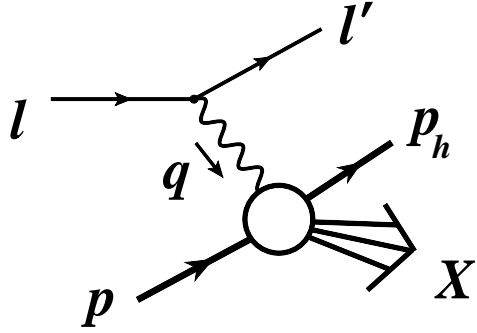}
\caption{Semi-inclusive deep-inelastic lepton--nucleon scattering
	with production of a final state hadron $h$.
	The external momenta of the incident ($\ell$) and scattered
	($\ell'$) leptons, virtual photon ($q$), target nucleon ($p$)
	and observed hadron ($p_h$) are labeled explicitly. 
        The unobserved final state hadrons are labelled by $X$.}
\label{fig:sidis}
\end{figure}

The semi-inclusive lepton--nucleon scattering process is illustrated
in Fig.~\ref{fig:sidis}, where the incident lepton (with momentum
$\ell$) scatters from an initial state nucleon ($p$) to a recoil
lepton ($\ell'$) via the exchange of a virtual photon ($q$),
producing a final state hadron $h$ (with momentum $p_h$).
In this section we first review the external kinematics and choice
of variables, before outlining the collinear factorization framework
for describing the hard scattering process.
After defining the hadronic tensor in terms of quark--nucleon and
quark--hadron correlation functions, we derive expressions
for the spin-averaged and spin-dependent cross sections in terms of
parton distribution and fragmentation functions at leading order
in the strong coupling constant and for finite values of $Q^2$.

\subsection{External kinematics}
\label{ssec:external}

We expand the four-momenta of the external particles in terms
of coplanar light-cone unit vectors $n$ and $\nbar$, satisfying
$n^2 = \nbar^2 = 0$ and $n \cdot \nbar=1$ \cite{EFP}.
The ``plus'' and ``minus'' components of a four-vector $v^\mu$
are defined by
$v^+ = v \cdot n     = (v^0 + v^3)/\sqrt{2}$ and
$v^- = v \cdot \nbar = (v^0 - v^3)/\sqrt{2}$.
We work in a class of reference frames where the initial nucleon
and virtual photon momenta are coplanar, so that
\begin{eqnarray}
p^\mu &=& p^+\, \nbar^\mu
       + \frac{M^2}{2 p^+}\, n^\mu,		\label{eq:p} \\
q^\mu &=& - \xi p^+\, \nbar^\mu 
       + \frac{Q^2}{2\xi p^+}\, n^\mu,		\label{eq:q}
\end{eqnarray}
where $M$ is the nucleon mass, $Q^2 = -q^2$, and the scaling
variable $\xi = - q^+/p^+$.
%
%
The $(p,q)$ collinear frames are related to each other by a boost
of parameter $p^+$ and contain, in particular, the target rest frame
(where the nucleon $p^+$ component is $p^+ = M/\sqrt{2}$) and the
Breit frame ($p^+=Q/(\sqrt{2}\xi)$).  For other choices of reference
frames see Appendix~A of Ref.~\cite{AHM09}.
In the collinear $(p,q)$ frame, $\xi$ is identical to the Nachtmann
scaling variable \cite{Nachtmann73, Greenberg, Frampton76},
\begin{equation}
\xi = \frac{2 x_B}{1 + \sqrt{1 + 4 x_B^2 M^2/Q^2}},
\label{eq:xi}
\end{equation}
where $x_B=Q^2/2p\cdot q$ is the Bjorken scaling variable, and 
in the Bjorken limit ($Q^2$ and $q^- \to \infty$ with $x_B$ fixed)
one finds that $\xi \to x_B$.
Momentum and baryon number conservation (and, for $K$ production,
strangeness conservation) impose an upper limit on $x_B$,
$x_B \leq x_B^{\rm max}$, where
\begin{equation}
\frac{1}{x_B^{\rm max}}
= 1 + \frac{m_h (m_h + 2 M_b) + M_b^2 - M^2}{Q^2},
\label{eq:xBlim}
\end{equation}
with $M_b = M$ for $\pi$ production and $M_b = M_\Lambda$
(the lightest hyperon mass) for $K$ production.
This value of $x_B$ corresponds to a final state consisting of
the nucleon and hadron $h$ at rest in the hadron's rest frame.

For spin-dependent scattering, the polarization vector $S^\mu$
of the initial state nucleon can be parametrized as
\begin{equation}
S^\mu = \frac{S_L}{M}
	\Big( p^+\, \nbar^\mu - \frac{M^2}{2 p^+}\, n^\mu + S_T^\mu
	\Big),
\label{eq:S}
\end{equation}
and satisfies the conditions $p \cdot S = 0$ and $S^2 = -1$.
In the case of a longitudinally polarized initial state nucleon
(which we consider in this work), one has $S_L = \pm 1$ and the
transverse spin vector $S_T^\mu = 0$.

The incident and scattered lepton momenta can be decomposed as
\begin{eqnarray}
\ell^\mu &=& \eta p^+\, \nbar^\mu
       + \Big( 1+\frac{\eta}{\xi} \Big)
	 \frac{Q^2}{2\xi p^+}\, n^\mu
       + \ell_\perp^{\,\mu},			\label{eq:l} \\
\ell'^\mu &=& (\eta + \xi)\, p^+\, \nbar^\mu
        + \frac{\eta}{\xi}\frac{Q^2}{2\xi p^+}\, n^\mu
        + \ell_\perp^{\,\mu},			\label{eq:l'}
\end{eqnarray}
where $\ell_\perp^\mu$ is the lepton transverse momentum four-vector,
$\eta = \ell^+/p^+$ is the lepton momentum fraction, and we assume
massless leptons, $\ell^2 = \ell'^2 = 0$.
After some algebra one can show that
\begin{equation}
\eta = \frac{\xi}{2y(1+\gamma^2)}
       \Big[ (2-y)\sqrt{1+\gamma^2} - y(1+\gamma^2) \Big],
\label{eq:eta}
\end{equation}
where $y = q \cdot p / \ell \cdot p$ and $\gamma^2 = 4 x_B^2 M^2/Q^2$.
In the target rest frame $y = \nu/E$ is the fractional energy
transfer from the lepton to the target, with $\nu$ and $E$ the
virtual photon and incident lepton energies, respectively.
The magnitude of the lepton transverse momentum is set by
four-momentum conservation,
\begin{equation}
\ell^2_\perp = - \bm{\ell}^2_\perp
	  = - \frac{\eta}{\xi}
	    \Big( 1+\frac{\eta}{\xi} \Big) Q^2.
\label{eq:ltsq}
\end{equation}

For the hadron produced in the final state, the momentum is
parametrized as
\begin{eqnarray}
p_h^\mu &=& \frac{\xi m_{h\perp}^2}{\zeta_h Q^2}\, p^+\, \nbar^\mu
	 + \frac{\zeta_h Q^2}{2 \xi p^+}\, n^\mu 
	 + p_{h\perp}^{\,\mu},			\label{eq:ph}
\end{eqnarray}
where $\zeta_h = p_h^-/q^-$ is the scaling fragmentation variable,
and the hadron transverse momentum four-vector $p_{h\perp}^{\,\mu}$
satisfies
	$p_{h\perp} \cdot n = p_{h\perp} \cdot \nbar = 0$,
with norm
	$p_{h\perp}^2 = -\bm{p}_{h\perp}^{\,2}$.
The squared transverse mass of the produced hadron $h$ is defined
by $m_{h\perp}^2 = m_h^2 + \bm{p}_{h\perp}^{\,2}$, where $m_h$ is
the mass of the hadron.
The variable $\zeta_h$ can be related to the fragmentation invariant
\begin{equation}
z_h = \frac{p_h \cdot p}{q \cdot p}
    = \frac{x_B}{\xi}
      \left( \zeta_h
	   + \frac{\xi^2}{\zeta_h} \frac{M^2 m_{h\perp}^2}{Q^4}
      \right).
\label{eq:z_h}
\end{equation}
In the target rest frame it coincides with the ratio of the produced
hadron to virtual photon energies, $z_h = E_h/\nu$, which is
frequently used in experimental analysis of SIDIS data.
In the Bjorken limit the fragmentation variable $\zeta_h \to z_h$,
while at finite $Q^2$ one has
\begin{eqnarray}
\zeta_h
&=& \frac{z_h}{2} \frac{\xi}{x_B}
    \left( 1 + \sqrt{1 - \frac{4 x_B^2 M^2 m_{h\perp}^2}{z_h^2\ Q^4}} 
    \right).
\label{eq:zeta_h}
\end{eqnarray}
Since the produced hadron's energy is bounded from below by
$E_h \geq m_{h\perp}$, one can show that $z_h \geq z_h^{\rm min}$,
where
\begin{equation}
z_h^{\rm min} = 2 x_B \frac{M m_h}{Q^2}.
\label{eq:zh_min}
\end{equation}
Combining Eqs.~(\ref{eq:zeta_h}) and (\ref{eq:zh_min}), one can show
that the corresponding minimum value of $\zeta_h$ is given by
$\zeta_h^{\rm min} = \xi M m_h / Q^2$.
In the target rest frame, $z_h^{\rm min}$ corresponds to the hadron
$h$ produced at rest, with the remaining final state hadrons moving
collectively in the direction of the virtual photon.
At the other extreme, conservation of four-momentum, baryon number,
and (for $K$ production) strangeness impose the upper limit
$z_h \leq z_h^{\rm max}$, where
\begin{align}
z_h^{\rm max} = 1 - 2x_B \frac{M (M_b-M)}{Q^2},
\label{eq:zh_max}
\end{align}
with again $M_b = M$ for $h = \pi$ and $M_b = M_\Lambda$ for $h = K$.
This limit corresponds to diffractive production of the observed
hadron with maximal energy.  As $Q^2 \to \infty$, both the upper
and lower limits become independent of $Q^2$,
$z_h^{\rm min} \to 0$ and $z_h^{\rm max} \to 1$.

As an alternative to the fragmentation invariant $z_h$, one can
define the invariant momentum fraction
\begin{equation}
z_e = \frac{2p_h \cdot q}{q^2}
    = \zeta_h - \frac{m_{h\perp}^2}{\zeta_h Q^2},
\label{eq:z_e}
\end{equation}
which is used in the study of hadron production in $e^+ e^-$ collisions.
The choice of this variable avoids mixing inclusive ($x_B$) and
semi-inclusive ($z_h$) variables as in Eq.~\eqref{eq:z_h}.
It also allows a clean separation of the current ($z_e > 0$) and
target ($z_e < 0$) fragmentation regions in the Breit frame, in which
$z_e = p_h^z/q^z$ is the ratio of the longitudinal components of the
hadron and photon momenta.
In the current region, where the observed hadrons are produced
with longitudinal momentum in the direction of the virtual photon,
$\zeta_h$ can be written in terms of the $z_e$ variable as
\begin{equation}
\zeta_h = \frac{z_e}{2} 
  \left( 1 + \sqrt{1 + \frac{4 m_{h\perp}^2}{z_e^2\, Q^2}}
  \right).
\label{eq:eta_h2}
\end{equation}
Hadrons produced in the current region have
$\zeta_h > \zeta_h^{(0)} \equiv \zeta_h(z_e=0)$, where
\begin{align}
\zeta_h^{(0)} = \frac{m_{h\perp}}{Q}.
\end{align}
Note that $\zeta_h(z_h^{\rm min}) \leq \zeta_h^{(0)}$, which reflects
the fact that a hadron produced at rest in the target rest frame
belongs to the target region.
Finally, in the Bjorken limit all three fragmentation variables
become equivalent, $\zeta_h \to z_h \to z_e$, and the current region
extends down to the smallest values of $z_h$, $\zeta_h^{(0)} \to 0$.

\begin{figure}[t]
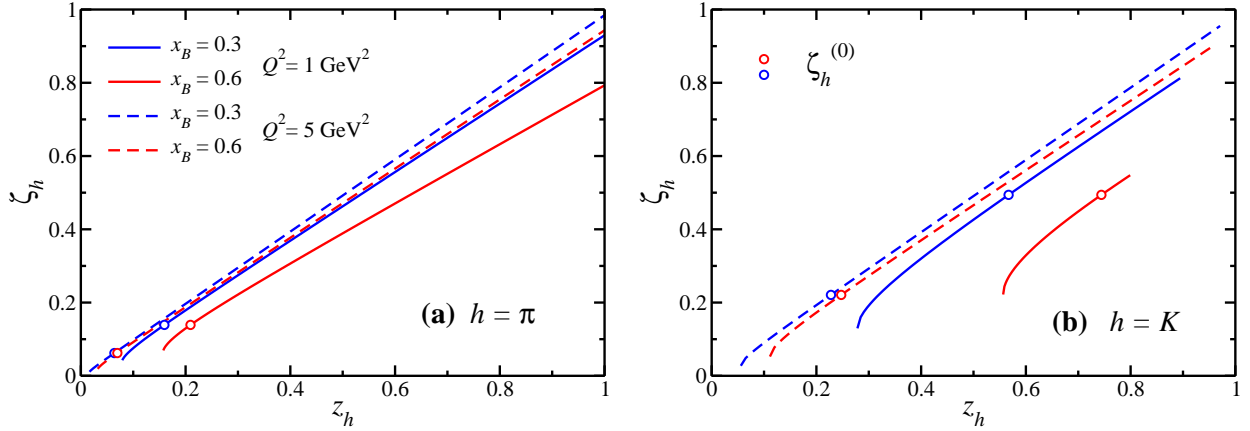

\includegraphics[width=8cm]{hmc_zetapi.eps}\ \ \
\includegraphics[width=8cm]{hmc_zetaK.eps}
\caption{Finite-$Q^2$ fragmentation variable $\zeta_h$ versus
	$z_h$ for the semi-inclusive production of
	(a) pions, $h = \pi$ and (b) kaons, $h = K$, at fixed
	values of $x_B = 0.3$ (blue curves) and 0.6 (red curves)
	for $Q^2 = 1$ (solid curves) and 5~GeV$^2$ (dashed curves).
	The curves are shown only in the kinematically allowed
	$z_h$ regions, and the boundaries between the
	current	($\zeta_h > \zeta_h^{(0)}$) and
	target ($\zeta_h < \zeta_h^{(0)}$) fragmentation
	regions are indicated by the open circles.}
\label{fig:zeta}
\end{figure}

The relation between the variable $\zeta_h$ and $z_h$ is illustrated
in Fig.~\ref{fig:zeta} for several fixed values of $x_B$ and $Q^2$.
At finite $Q^2$ the kinematically allowed regions of $z_h$ are
determined by Eqs.~(\ref{eq:zh_min}) and (\ref{eq:zh_max}), and the
boundaries between the current and target fragmentation regions occur
at $\zeta_h = \zeta_h^{(0)}$.
For the production of pions, at low $x_B \lesssim 0.3$ and high
$Q^2 \gtrsim 5$~GeV$^2$ the differences between the two variables
are almost negligible, and begin to be noticeable only for the
highest $z_h$ values at lower $Q^2$ or higher $x_B$.
At these kinematics, pions are produced in the current region
for $z_h \gtrsim 0.08-0.15$.
At high $x_B \approx 0.6$ and low $Q^2 \approx 1$~GeV$^2$, the effects
are more pronounced, with deviations of $\sim 30\%$ as $z_h \to 1$,
and current fragmentation begins at a slightly higher $z_h$.
The effects of the kinematic lower limit in $z_h$
[Eq.~(\ref{eq:zh_min})] is noticeable only at low $Q^2$ and high $x_B$.

For kaons, the effects at high $Q^2$ and low $x_B$ are again
negligible, although the larger $K$ mass enhances the differences
relative to the pion at the same kinematics.
In particular, at low $Q^2 = 1$~GeV$^2$ the lower limit on
$\zeta_h$ is dramatically increased, and the current region
is pushed to higher values of $z_h$.
At sufficiently large $x_B$, the phase space for $K$ production
eventually vanishes; for $Q^2=1$~GeV$^2$, for example, no kaons
can be produced in the current fragmentation region with
$x_B \gtrsim 0.66$, and no kaons can be produced at all for
$x_B \gtrsim 0.8$.

\subsection{Collinear factorization}
\label{ssec:cf}

\begin{figure}[t]
\includegraphics[height=8cm]{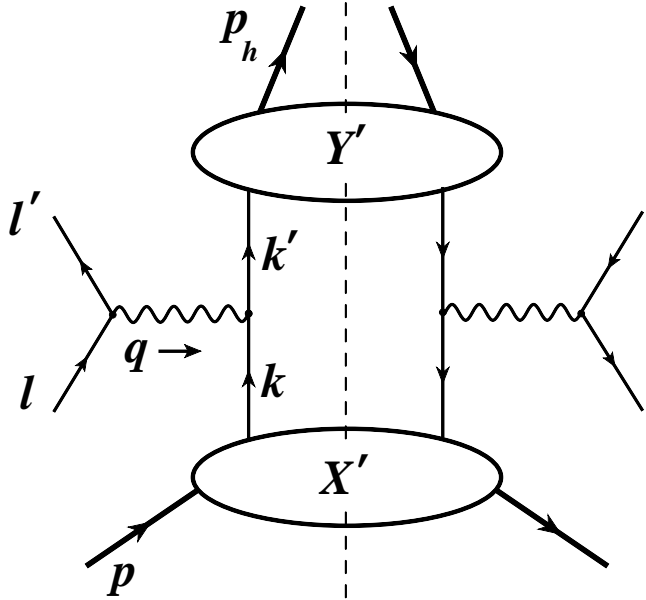}
\caption{Semi-inclusive deep-inelastic lepton--nucleon scattering
	with production of a final state hadron $h$ at leading order
	in $\alpha_s$.
	The internal momenta of the initial ($k$) and scattered
	quarks ($k'$) are labeled explcitly.  
        The intermediate state $X'$ represents a
	nucleon with a quark removed, and the state $Y'$ results
	from the fragmenting parton with the hadron $h$ removed.
	The dashed vertical line represents the cut in the forward
	scattering amplitude.}
\label{fig:sidis-pQCD}
\end{figure}

At leading order in the strong coupling $\alpha_s$, the SIDIS reaction
proceeds through the hard scattering of the virtual photon from an
initial state quark with momentum $k$ to a quark with momentum
$k' = k+q$, which then fragments to a hadron $h$, as illustrated in
Fig.~\ref{fig:sidis-pQCD}.
Higher order processes involving gluon radiation and scattering from
$q\bar q$ pairs can be considered, but for clarity of the derivation
of the finite-$Q^2$ corrections we restrict ourselves to the leading
order calculation.

The parton four-momenta can be parametrized, in analogy with the
external variables in Sec.~\ref{ssec:external}, in terms of the
light-cone vectors $n$ and $\nbar$ as
\begin{eqnarray}
k^\mu &=& xp^+\, \bar{n}^\mu
       + \frac{k^2 + \bm{k}_\perp^2}{2 x p^+}\, n^\mu
       + k^\mu_\perp,				\label{eq:k} \\
k'^\mu &=& \frac{k'^2 + \bm{k}'^2_\perp}{2 p_h^-/z}\, \bar{n}^\mu
       + \frac{p_h^-}{z}\, n^\mu
       + k'^\mu_\perp,				\label{eq:k'}
\end{eqnarray}
where $x = k^+ / p^+$ is the light-cone momentum fraction of the
nucleon carried by the struck quark, and $z = p_h^-/k'^-$ is the 
light-cone fraction of the fragmenting quark carried by the
hadron $h$.
The parton transverse momentum four-vectors $k_\perp$
and $k'_\perp$ are orthogonal to $n$ and $\nbar$,
$k_\perp \cdot n = k_\perp \cdot \bar n = 0$ and
$k'_\perp \cdot n = k'_\perp \cdot \bar n = 0$,
with norms $k^2_\perp = -\bm{k}^2_\perp$ and
$k'^2_\perp = -\bm{k}'^2_\perp$.

The essence of the collinear factorization approach is to
expand the hard scattering term around ``collinear'' momenta
$\widetilde k$ and $\widetilde k'$,
\begin{eqnarray}
\widetilde k^\mu
&=& xp^+\, \bar{n}^\mu
 + \frac{\widetilde k^2}{2 x p^+}\, n^\mu,	\label{eq:kon} \\
\widetilde k'^\mu
&=& \frac{\widetilde k'^2 + {\bm p}_{h\perp}^2/z^2}
	 {2 p_h^-/z}\, \bar{n}^\mu
 + \frac{p_h^-}{z}\, n^\mu
 + \frac{p_{h\perp}^\mu}{z},			\label{eq:k'on}
\end{eqnarray}
where the initial and final parton virtualities
$\widetilde k^2$ and $\widetilde k'^2$ are kept for generality.
In this approximation the transverse momentum of the initial parton
is neglected, and the transverse momentum of the final parton is
taken along the direction of the hadron $h$, making these collinear
to the proton target and produced hadron, respectively.

Defining the partonic analog of the Bjorken scaling variable $x_B$
at finite $Q^2$ by
\begin{equation}
\hat x
= -\frac{q^2}{2\widetilde k\cdot q}
=  \frac{\xi}{x} \, \frac{1}{1-\xi^2 \widetilde k^2 / x^2 Q^2},
\label{eq:xf}
\end{equation}
one can show that $\hat x$ lies within the range
$\hat x_{\rm min} \leq \hat x \leq \hat x_{\rm max}$,
where \cite{AQ08, AHM09}
\begin{equation}
\frac{1}{\hat x_{\rm min}}
= \frac{1}{x_B} - \frac{2 M m_h + \widetilde k^2}{Q^2},\ \ \ \ \ \
\frac{1}{\hat x_{\rm max}}
= 1 + \frac{m_h^2}{\zeta_h Q^2}
    - \frac{\widetilde k^2}{Q^2}
      \Big( 1 - \frac{\xi m_h^2}{x \zeta_h Q^2} \Big).
\label{eq:xflimits}
\end{equation}
Here the lower limit $\hat x_{\rm min}$ corresponds to the minimal
mass of collinear spectator partons, while the upper limit
$\hat x_{\rm max}$ arises from the minimum value of the current
jet invariant mass.
These limits are consistent with the limit on $x_B$ in
Eq.~(\ref{eq:xBlim}) for any $\widetilde k^2 \geq x (\zeta_h-1)Q^2/\xi$,
and in the Bjorken limit the range of $\hat x$ reduces to being between
$x_B$ and 1, as required.
Analogous limits can be found for the partonic fragmentation
variable $z$, $z_{\rm min} \leq z \leq 1$, where
\begin{equation}
\frac{1}{z_{\rm min}}
= \frac{1}{\zeta_h}
  \Big( 1 + \frac{\xi}{x} \frac{\widetilde k^2}{Q^2} \Big).
\label{eq:wlimits}
\end{equation}

For the practical implementation of the finite-$Q^2$ kinematical
limits, a choice of the initial and scattered parton virtualities
$\widetilde k^2$ and $\widetilde k'^2$ must be made.
For the initial parton the requirement that the collinear parton
virtuality is independent of $x$ leads to the restriction
$\widetilde k^2 \geq 0$.
For light, bound initial-state partons ($\widetilde k^2 \leq 0$) 
this constrains therefore the collinear expansion to be around
$\widetilde k^2 = 0$.
Determining the virtuality $\widetilde k'^2$ of the scattered
parton is generally less clear, on the other hand.
For the leading order hard scattering in Fig.~\ref{fig:sidis-pQCD},
conservation of four-momentum and the condition $\widetilde k^2 = 0$
constrain the parton momentum fraction $x$ to equal $\xi_h$, where
\begin{equation}
\xi_h = \xi \Big( 1 + \frac{\widetilde k'^2}{Q^2} \Big).
\label{eq:xi_h}
\end{equation}
In order for $x$ to respect the limits in Eq.~(\ref{eq:xflimits}),
the choice $\widetilde k'^2 = m_h^2/\zeta_h$ was proposed in
Ref.~\cite{AHM09}, giving $\xi_h = \xi (1 + m_h^2/\zeta_h Q^2)$.
Larger $\widetilde k'^2$ values would also allow $\hat x$ to fall
within the bounds in Eq.~(\ref{eq:xflimits}); however, the above
choice is the closest to the physical quark mass, and is the one
adopted in our numerical analysis here.  The dependence of the
calculated cross sections on the choice of $\widetilde k'^2$ is
explored further in Sec.~\ref{sec:results}.

\subsection{SIDIS at finite $Q^2$}
\label{ssec:cross}

The cross section for the SIDIS process can be written as a product
of hadronic ($W^{\mu \nu}$) and leptonic ($L_{\mu\nu}$) tensors
\cite{Bacchetta07},
\begin{equation}
\frac{d\sigma}{dx_B\, dy\, d^3\bm{p}_h/2E_h}
= \frac{\pi \alpha^2 y}{Q^4}\,
  2 M W^{\mu\nu}\, L_{\mu\nu},
\label{eq:dsig}
\end{equation}
where $\alpha$ is the electromagnetic fine structure constant.
The leptonic tensor can be computed straightforwardly from QED,
\begin{equation}
L_{\mu\nu}(\ell,\ell',\lambda)
= 2 \ell_\mu \ell'_\nu + 2 \ell_\nu \ell'_\mu + q^2\, g_{\mu \nu}
+ 2 i \lambda\, \epsilon_{\mu \nu \rho \sigma}\,
  \ell^\rho \ell'^\sigma,
\label{eq:Lmunu}
\end{equation}
where $\lambda$ is the lepton helicity.

The semi-inclusive hadronic tensor is defined in terms of matrix
elements of the electromagnetic current operator $J^\mu$ between
the initial state nucleon with spin $S$ and the final state with
a hadron $h$ and unobserved hadrons $X$,
\begin{eqnarray}
2M W^{\mu\nu}(p,S,q,p_h)
&=& \frac{1}{(2 \pi)^3}
  \sum_X \int \frac{ d^3\bm{p}_X} {2 E_X}
  \delta^{(4)} \big( p + q - p_X - p_h \big)	\nonumber\\
& & \hspace*{2cm} \times
  \langle N,S  | J^\mu(0) | h, X \rangle
  \langle h, X | J^\nu(0) | N,S  \rangle,
\end{eqnarray}
where we use the shorthand notation
$d^3\bm{p}_X / 2E_X = \prod_{i\in X} d^3\bm{p}_i / 2E_i$,
with $p_X$ the total momentum of teh unobserved hadrons.
The hadron tensor can be expressed at leading order in $\alpha_s$
in terms of quark-quark correlators $\Phi_q$ and $\Delta_q^h$,
associated with the quark distribution and fragmentation functions,
respectively \cite{CSS88, CS82, Mulders01},
\begin{equation}
2M W^{\mu\nu}(p,S,q,p_h)
= \sum_q e_q^2 \int d^4k\ d^4k'\
  \delta^{(4)}(\widetilde k + q - \widetilde k')\,
  \Tr \left[ \Phi_q(p,S,k)\, \gamma^\mu\, \Delta_q^h(k',p_h)\,
	     \gamma^\nu
      \right],
\label{eq:Wmunu}
\end{equation}
where the sum is taken over quark and antiquark flavors $q$.
Note that since the parton momenta in the $\delta$-function have
been  approximated by their collinear components, the integrations
over $dk^-\, d^2\bm{k}_{\perp}$ and $dk'^+\, d^2\bm{k}'_\perp$ act
directly on the correlators $\Phi_q$ and $\Delta_q^h$.

The correlator associated with the parton distribution function
is defined, in the light-cone gauge, as \cite{Bacchetta12}
\begin{equation}
\Phi_q(p,S,k)
= \frac{1}{(2 \pi)^3}
  \sum_{X'} \int \frac{d^3\bm{p}_{X'}}{2 E_{X'}}
  \delta^{(4)} \big( p - k - p_{X'} \big)
  \langle N,S | \bar{\psi}_q(0) | X'  \rangle
  \langle X'  | \psi_q(0)       | N,S \rangle,
\label{eq:Phi_def}
\end{equation}
where $\psi_q$ is the quark field operator, and $E_{X'}$ and $p_{X'}$
are the energy and momenta of the intermediate state $X'$ corresponding
to a nucleon with a quark removed.
Similarly, for the quark fragmentation correlator one has
\cite{Bacchetta12}
\begin{equation}
\Delta_q^h(k',p_h)
= \frac{1}{(2 \pi)^3}
  \sum_{Y'} \int \frac{d^3\bm{p}_{Y'}}{2 E_{Y'}}
  \delta^{(4)} \big( k - p_h - p_{Y'} \big)
  \langle 0     | \psi_q(0)       | h, Y' \rangle
  \langle h, Y' | \bar{\psi}_q(0) | 0     \rangle,
\label{eq:Delta_def}
\end{equation}
where $E_{Y'}$ and $p_{Y'}$ are the energy and momenta of the state
$Y'$ resulting from the fragmenting quark with the hadron $h$ removed.

The leading twist contributions to the SIDIS cross section can be
extracted by expanding the integrated correlators and selecting
the terms linear in the light-cone vectors.
For the parton distribution function, one finds
\begin{equation}
\Phi_q(x)
\equiv \int dk^- d^2\bm{k}_\perp\, \Phi_q(p,S,k)\,
=\, \frac12\, q(x)\, \nbslash\,
+\, \frac12 S_L\, \Delta q(x)\, \gamma^5\, \nbslash\, +\, \ldots,
\label{eq:Phi_exp}
\end{equation}
where the first and second terms define the spin-averaged, $q(x)$,
and spin-dependent, $\Delta q(x)$, distribution functions, and the
ellipsis indicates higher twist contributions \cite{Bacchetta07}.
The fragmentation function $D_q^h(z)$ is analogously defined from
the $\Delta_q^h$ correlator by
\begin{equation}
\Delta_q^h(z)
\equiv \frac{z}{2} \int dk'^+ d^2\bm{k}'_\perp\, \Delta_q^h(k',p_h)\,
=\, \frac12 D_q^h(z)\, \nslash\, +\, \ldots
\label{eq:Delta_exp}
\end{equation}
Inverting Eqs.~(\ref{eq:Phi_exp})--(\ref{eq:Delta_exp}), one can
write the quark distribution and fragmentation functions explicitly
by projecting with the appropriate Dirac operators,
\begin{align}
q(x)
&= \frac12 \int dk^- d^2\bm{k}_\perp\,
   \Tr \left[ \gamma^+ \Phi_q(p,S,k) \right]_{k^+=xp^+}
							\nonumber\\
&= \frac{1}{4\pi} \int dw^-\, e^{i xp^+ w^-}
   \langle N | \overline{\psi}_q(0)\, \gamma^+\, \psi_q(w^- n)
   | N \rangle,						\\
\Delta q(x)
&= \frac12 \int dk^- d^2\bm{k}_\perp\,
   \Tr \left[ \gamma^5 \gamma^+ \Phi_q(p,S,k) \right]_{k^+=xp^+}
							\nonumber\\
&= \frac{1}{4\pi} \int dw^-\, e^{i xp^+ w^-}
   \langle N | \overline{\psi}_q(0)\, \gamma^5\, \gamma^+\, \psi_q(w^- n)
   | N \rangle,						\\ 
D_q^h(z)
&= \frac{z}{4} \int  dk'^+ d^2\bm{k}'_\perp\,
   \Tr \left[ \gamma^- \Delta_q^h(k',p_h) \right]_{k'^-=p_h^-/z}
							\nonumber\\
&= \frac{z}{8\pi} \sum_{Y'} \int dw^+\, e^{i(p_h^-/z)w^+}
   \langle 0     | \psi_q(w^+ \nbar)                |h, Y' \rangle
   \langle h, Y' | \overline{\psi}_q(0) \gamma^-| 0    \rangle, 
\end{align}
where $\omega^\pm$ are light-cone coordinates.
(For ease of notation we omit the $Q^2$ dependence in the arguments
of the quark distribution and fragmentation functions.)
The fragmentation function $D_q^h(z)$ here is defined with the
standard normalization,
$\sum_h \int_0^1 dz\, z\, D_q^h(z) = 1$ \cite{Mulders01}.

To compute the hadronic tensor in Eq.~(\ref{eq:Wmunu}), we can
decompose the $\delta^{(4)}(\widetilde k + q - \widetilde k')$
function along the $+$, $-$ and transverse components of the
momenta.  The $\delta$-functions for the $+$ and $-$ components
constrain the partonic variables to $x = \xi_h$ and $z = \zeta_h$,
respectively.  The $\delta$-function for the transverse component
forces the transverse momentum of the produced hadron $h$ to vanish,
$p_{h\perp} = z\, k'_\perp = 0$.
Nonzero transverse momentum hadrons can be produced via higher order
perturbative QCD processes, or from intrinsic transverse momentum
in the parton distribution functions themselves \cite{Bacchetta07}.
With these constraints, the hadron tensor can then be factorized
into products of parton distribution and fragmentation functions
evaluated at $\xi_h$ and $\zeta_h$, respectively,
\begin{eqnarray}
2M W^{\mu\nu}(p,S,q,p_h)
&=& \frac{\zeta_h}{2} \sum_q e_q^2\,
  \delta^{(2)}(\bm{p}_\perp)
  \Big( \Tr \left[ \nbslash \gamma^\mu \nslash \gamma^\nu
            \right] q(\xi_h)				\nonumber\\
& & \hspace*{2.4cm}
  +\ S_L \Tr \left[ \gamma^5 \nbslash \gamma^\mu \nslash \gamma^\nu
	    \right] \Delta q(\xi_h)
  + \cdots
  \Big) D_q^h(\zeta_h).
\label{eq:TMC_hadtens}
\end{eqnarray}
The main effect of the hadron masses at finite kinematics is
therefore a replacement of the Bjorken limit scaling variables
$x_B$ and $z_h$ by their finite-$Q^2$ analogs.
However, since $\xi_h$ depends explicitly on $m_h$, and $\zeta_h$
depends on $z_h$ and $x_B$, the scattering and fragmentation parts
of the hadronic tensor at finite $Q^2$ are not independent.

Contracting the hadronic tensor in Eq.~(\ref{eq:TMC_hadtens}) with
the leptonic tensor in Eq.~(\ref{eq:Lmunu}) enables the leading order
spin-averaged ($\sigma_h$) and spin-dependent ($\Delta\sigma_h$)
SIDIS cross sections to be written in terms of parton distributions
evaluated at the new scaling variables,
\begin{subequations}
\begin{eqnarray}
\sigma_h\ \equiv\
\frac{1}{2} \frac{d\sigma_h^{\uparrow\uparrow+\downarrow\uparrow}}
     {dx_B\, dQ^2\, dz_h}
&=& \frac{2\pi \alpha^2}{Q^4}
    \frac{y^2}{1-\varepsilon}
    \bar{\sigma}_h,
\label{eq:dsig_unp}			\\
\Delta\sigma_h\ \equiv\
\frac{d\sigma_h^{\uparrow\uparrow-\downarrow\uparrow}}
     {dx_B\, dQ^2\, dz_h}
&=& \frac{4\pi \alpha^2}{Q^4}
    \frac{y^2 \sqrt{1-\varepsilon^2}}{1-\varepsilon}
    \Delta\bar{\sigma}_h,
\label{eq:dsig_pol}
\end{eqnarray}
\label{eq:dsig_both}
\end{subequations}
where the reduced unpolarized and polarized cross sections are
defined as
\begin{subequations}
\begin{eqnarray} 
\bar{\sigma}_h
&=& J_h \sum_q e_q^2\, q(\xi_h,Q^2)\, D_q^h(\zeta_h,Q^2),	\\
\Delta\bar{\sigma}_h
&=& J_h \sum_q e_q^2\, \Delta q(\xi_h,Q^2)\, D_q^h(\zeta_h,Q^2).
\end{eqnarray} 
\label{eq:dsigbar}
\end{subequations}%
In Eqs.~(\ref{eq:dsig_both}) the arrows denote the spins of the
lepton and target nucleon, and the dependence of the functions
on the scale $Q^2$ is made explicit.
In Eqs.~(\ref{eq:dsig_both}) the variable
\begin{equation}
\varepsilon
= \frac{1 - y - y^2 \gamma^2/4}{1 - y + y^2 (1+\frac12\gamma^2)/2}
\end{equation}
is the ratio of longitudinal to transverse photon flux, and in
Eqs.~(\ref{eq:dsigbar}) $J_h$ is a scale dependent Jacobian factor,
$J_h = d\zeta_h/dz_h
	  = (1 - M^2\xi^2/Q^2)
          / (1-\xi^2 M^2 m_h^2/\zeta_h^2 Q^4)$,
with $J_h \to 1$ at large $Q^2$.

Note that at the maximum value of $x_B$ allowed for SIDIS
[see Eq.~(\ref{eq:xBlim})] the finite-$Q^2$ variable $\xi_h$
satisfies $\xi_h < \xi_h (x_B=x_B^{\rm max}) < 1$.
As in the case of inclusive DIS \cite{AQ08}, the SIDIS cross
section therefore does not vanish as $x_B \to x_B^{\rm max}$,
which reflects the well-known threshold problem in which the
leading twist structure function is nonzero for $x_B \geq 1$
\cite{DGP77, Tung79, Steffens06, Steffens12}.
Analogously, for the finite-$Q^2$ fragmentation varialble one
has $\zeta_h < \zeta_h (z_h=z_h^{\rm max}) < 1$, and since the
fragmentation function does not vanish as $z_h \to z_h^{\rm max}$,
the perturbatively calculated SIDIS cross section can also 
exceed the fragmentation threshold.

\begin{figure}[t]
\includegraphics[width=8.5cm]{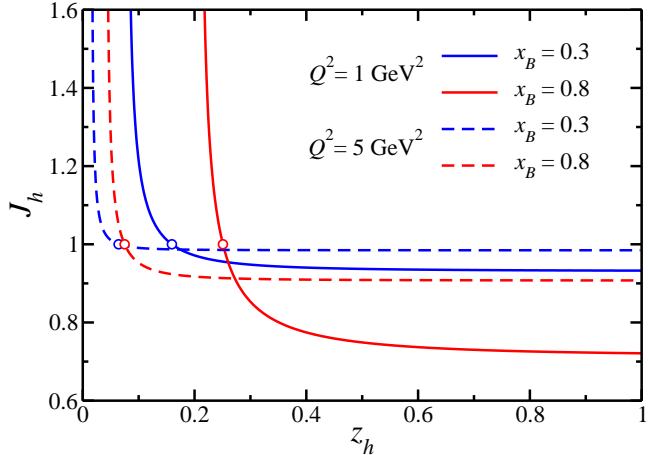}
\caption{Jacobian factor $J_h$ versus $z_h$ for pion production
	($h = \pi$) at fixed values of $x_B = 0.3$ (blue curves)
	and 0.8 (red curves), at $Q^2 = 1$ (solid curves) and
	5~GeV$^2$ (dashed curves).
	The open circles denote the boundary between the target
	(small $z_h$) and current (large $z_h$) fragmentation
	regions.}
\label{fig:Jac}
\end{figure}

Before exploring the dependence of the SIDIS cross sections on the
finite-$Q^2$ scaling variables in the next section, it is useful to
first establish the purely kinematic corrections, independent of the
PDFs and fragmentation functions, which augment the finite-$Q^2$
results from their scaling limit.  In Fig.~\ref{fig:Jac} we show the
$z_h$ dependence of the Jacobian factor $J_h$ in Eqs.~(\ref{eq:dsigbar})
at fixed values of $x_B$ and $Q^2$ for the case of pion production
($h = \pi$).
For $x_B = 0.3$ the factor $J_h$ deviates very little for unity
over most of the range of $z_h$, with an upturn only at small $z_h$,
$z_h \lesssim 0.1$ for $Q^2 \geq 1$~GeV$^2$.  At higher $x_B$ values
the finite-$Q^2$ effects are more visible, with $J_h$ spiking above
unity at $z_h \lesssim 0.2$ for $x_B = 0.8$ and $Q^2 = 1$~GeV$^2$,
and decreasing to $\approx 25\%$ below unity at large $z_h$.
This behavior should be kept in mind when assessing the numerical
effects of the HMCs in the cross sections in the next section.
One should also note that the upturn in $J_h$ is almost entirely
confined to the target fragmentation region (small $z_h$), where
the validity of calculations based on the perturbative handbag
diagram in Fig.~\ref{fig:sidis-pQCD} is more questionable, and
factorization in terms of fracture functions \cite{Trentadue93,
Graudenz94, deFlorian95} may be more appropriate.

\section{Phenomenological Implications}
\label{sec:results}

In this section we examine numerically the phenomenological consequences
of the finite-$Q^2$ rescaling of the SIDIS cross section derived in
Sec.~\ref{sec:semi-intro}, and explore their impact on future
hadron production experiments.

\subsection{Kinematical dependence of HMCs}
\label{ssec:kin}

\begin{figure}[t]
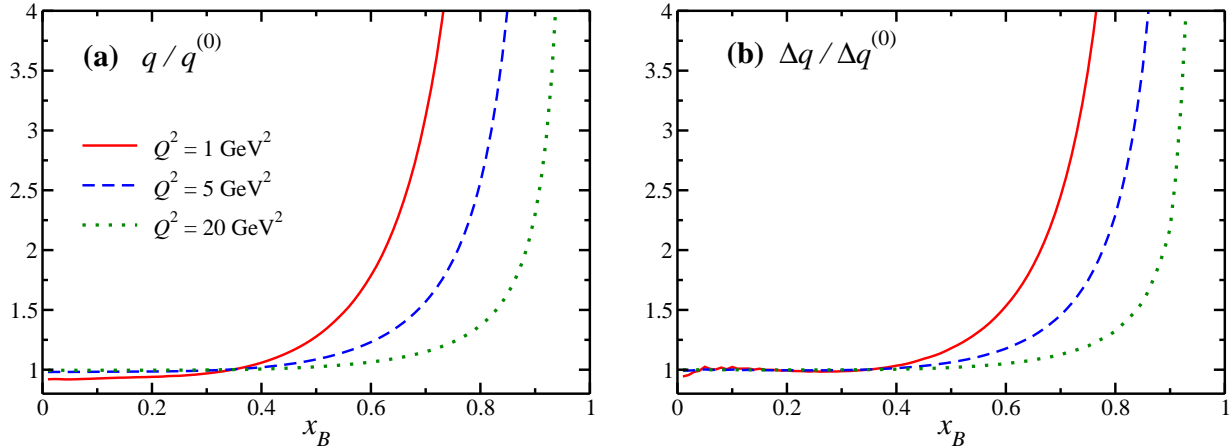

\includegraphics[width=7.8cm]{hmc_PDFunp.eps}\ \ \ \ \
\includegraphics[width=7.8cm]{hmc_PDFpol.eps}
\caption{Ratios of (a) spin-averaged $q = u + d$ and
	(b) spin-dependent $\Delta q = \Delta u + \Delta d$
	isoscalar PDFs to the corresponding massless limit
	distributions, $q^{(0)}$ and $\Delta q^{(0)}$,
	as a function of $x_B$ at various fixed $Q^2$.
	The finite-$Q^2$ scaling variable $\xi_h$ here is
	evaluated for $m_h = m_{\pi}$ and $\zeta_h = 0.2$.}
\label{fig:PDF}
\end{figure}

To disentangle the separate HMC effects in the SIDIS cross sections
arising from the PDFs and fragmentation functions, in Fig.~\ref{fig:PDF}
we illustrate the ratios of PDFs at finite $Q^2$ to the corresponding
massless limit distributions.  For a systematic comparison we consider
both the spin-averaged isoscalar $q=u+d$ and spin-dependent
$\Delta q = \Delta u + \Delta d$ distributions at several fixed
values of $Q^2$ from $Q^2=1$~GeV$^2$ to 20~GeV$^2$.
For the spin-averaged and spin-dependent PDFs we use the
leading order CT \cite{CT10} and LSS \cite{LSS05} parametrizations,
respectively, and evaluate the scaling variable $\xi_h$ for the case
of pion production, $m_h = m_\pi$, and for a typical value of  
$\zeta_h = 0.2$. The results for other $\zeta_h$ values are similar, 
essentially given by the PDF evaluated at a rescaled value of $Q^2$
[see Eq.~(\ref{eq:xi_h})].

The most dramatic feature in the ratios is the steep rise at large
$x_B$, which sets in at smaller $x_B$ values for decreasing $Q^2$.
The results are qualitatively similar for the unpolarized and
polarized distributions, with the rise delayed to slightly larger
$x_B$ for the latter.  The differences between the unpolarized and
polarized PDF ratios for the most part reflect the differences in the
shapes of the respective input PDFs.  
However, the qualitative features of the results do not change when
using other leading order distributions, such as the unpolarized
GJR \cite{GJR09} and polarized BB \cite{BB02} parametrizations.
Generally, the behavior of the ratios observed in Fig.~\ref{fig:PDF}
is reminiscent of that found in previous studies of TMCs for
inclusive DIS \cite{Schienbein08}.

\begin{figure}[t]
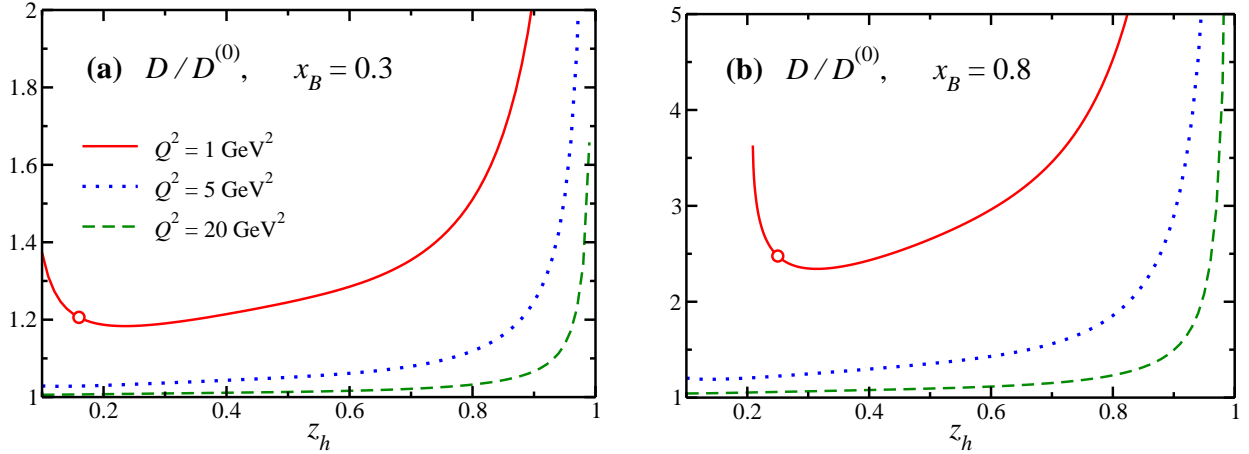

\includegraphics[width=7.9cm]{hmc_FFx3.eps}\ \ \ \ \ \ \
\includegraphics[width=7.6cm]{hmc_FFx8.eps}
\caption{Ratio of $\pi^+ + \pi^-$ isoscalar fragmentation functions
	$D$ to the corresponding massless limit functions $D^{(0)}$ at
	fixed $Q^2$ values for (a) $x_B = 0.3$ and (b) $x_B = 0.8$. 
	The open circles denote the boundary between the target
	and current fragmentation regions.}
\label{fig:FF}
\end{figure}

For the fragmentation functions, ratios of the finite-$Q^2$
isoscalar functions for $\pi^+ + \pi^-$ production to those in the
massless limit $D^{(0)}$ are displayed in Fig.~\ref{fig:FF} for
fixed values of $x_B$ and $Q^2$, using the leading order HKNS
parametrization \cite{HKNS07} for the fragmentation functions.
At $x_B=0.3$ the fragmentation function ratio at $Q^2=1$~GeV$^2$
is enhanced by $\approx 20-30\%$ for $z_h \lesssim 0.7$, before
rising rapidly as $z_h \to 1$.  The effect is less pronounced
with increasing $Q^2$, with a smaller enhancement of the ratio
and a delayed (though even more dramatic) rise at large $z_h$.
The fragmentation function ratios at $x_B = 0.8$ in
Fig.~\ref{fig:FF}(b), on the other hand, display a significantly
stronger enhancement, particularly at the lowest $Q^2$ value.
Here the effect is about an order of magnitude larger, and features
a striking upturn at $z_h \lesssim 0.2$, where the finite-$Q^2$
fragmentation function becomes several times larger than the
high-$Q^2$ limit.  As outlined in Ref.~\cite{AHM09}, this arises
from the shape of the fragmentation function at finite-$Q^2$
kinematics.
The general features of the results, however, remain unchanged
if one uses the KKP parametrization \cite{KKP00}, for instance.

In particular, expanding the hadron mass corrected fragmentation
function in a Taylor series about the massless limit, one can write
the corrected to uncorrected ratio as
\begin{equation}
\frac{D(\zeta_h)}{D(z_h)} \approx
1 + \left.\frac{dD(z_h)}{dz_h}\right|_{\zeta_h}
          \frac{(\zeta_h - z_h)}{D(z_h)}.
\label{eq:TMCdif}
\end{equation}
The $z_h$ dependence of the correction is thus determined by the
negative shift in the fragmentation variable ($\zeta_h-z_h$)
and by the $z_h$ slope of $D(z_h)$.  Since the pion fragmentation
function is generally a decreasing function of $z_h$ at small $z_h$,
the ratio is driven upward as $z_h \to z_h^{\rm min}$, where
$|\zeta_h-z_h|$ is maximum.  Note that for kaons and protons,
in contrast, the slope at small $z_h$ can be positive, which
would lead to a suppression of the mass corrected function at
$z_h \sim z_h^{\rm min}$.   
At the other extreme, in the exclusive production limit the
fragmentation function ratio becomes divergent for the same reason
as the PDFs; namely, the functions in the scaling limit vanish
as $x_B \to 1$ or $z_h \to 1$, whereas the finite-$Q^2$ scaling
variable and the corresponding rescaled functions remain finite.

\begin{figure}[t]
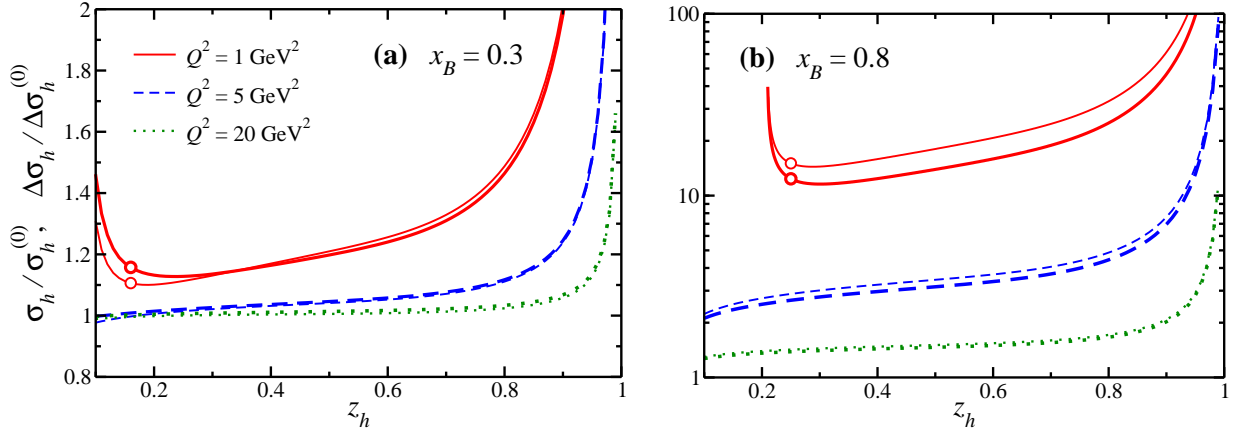

\includegraphics[width=8.2cm]{hmc_x3.eps}\ \ \ \ 
\includegraphics[width=7.5cm]{hmc_x8.eps}
\caption{Ratios of spin-averaged ($\sigma_h/\sigma_h^{(0)}$, thin lines)
	and spin-dependent ($\Delta\sigma_h/\Delta\sigma_h^{(0)}$, thick
	lines) cross sections with and without HMCs for semi-inclusive
	$\pi^+ + \pi^-$ production as a function of $z_h$,
	at fixed $Q^2$ for (a) $x_B = 0.3$ and (b) $x_B = 0.8$.
	Note the logarithmic scale on the ordinate in (b).  
	The open circles denote the boundaries between the target
	and current fragmentation regions.}
\label{fig:sig}
\end{figure}

Combining the effects of the HMCs in the parton distribution and
fragmentation functions, in Fig.~\ref{fig:sig} we show ratios of
the SIDIS spin-averaged and spin-dependent cross sections with
and without HMCs as a function of $z_h$, for several fixed values
of $Q^2$ and $x_B$.  Specifically, we consider scattering from a
proton target, with the production of $\pi^+ + \pi^-$ mesons in the
final state.  The massless limit cross sections $\sigma_h^{(0)}$
and $\Delta\sigma_h^{(0)}$ are defined by taking the high-$Q^2$
limits of the scaling variables in the arguments of the PDFs and
fragmentation functions,
$\sigma_h^{(0)} \equiv
 \sigma_h(\xi_h \to x_B, \zeta_h \to z_h)$
and
$\Delta\sigma_h^{(0)} \equiv
 \Delta\sigma_h(\xi_h \to x_B, \zeta_h \to z_h)$.

Overall, the $z_h$ dependence of the cross section ratios follows
the trends indicated in Figs.~\ref{fig:PDF} and \ref{fig:FF} for
the PDF and fragmentation function ratios, with strong enhancement
of the finite-$Q^2$ cross sections at large $z_h$, and decreasing
effects at higher $Q^2$.  At the lower $x_B$ value ($x_B=0.3$
in Fig.~\ref{fig:sig}(a)), the HMC effects in the spin-dependent
(thick lines) and spin-averaged (thin lines) ratios are very similar,
which reflects the qualitatively similar shapes of the $u$ and
$\Delta u$ distributions at intermediate $x_B$.  (Note that the
fragmentation functions for $\pi$ production are the same for the
spin-dependent and spin-averaged cross sections.)
While small differences are visible at $Q^2=1$~GeV$^2$, at the
higher $Q^2$ values the unpolarized and polarized ratios are almost
indistinguishable.
The differences are more striking at larger $x_B$ ($x_B=0.8$ in
Fig.~\ref{fig:sig}(b)), where the effects on the spin-averaged
cross section are somewhat larger than on the spin-dependent
cross section.
This stems directly from the delayed rise above unity of the
$\Delta q/\Delta q^{(0)}$ PDF ratio in Fig.~\ref{fig:PDF}(b)
at high $x_B$ values compared with the corresponding $q/q^{(0)}$
ratio in Fig.~\ref{fig:PDF}(a).

\begin{figure}[t]
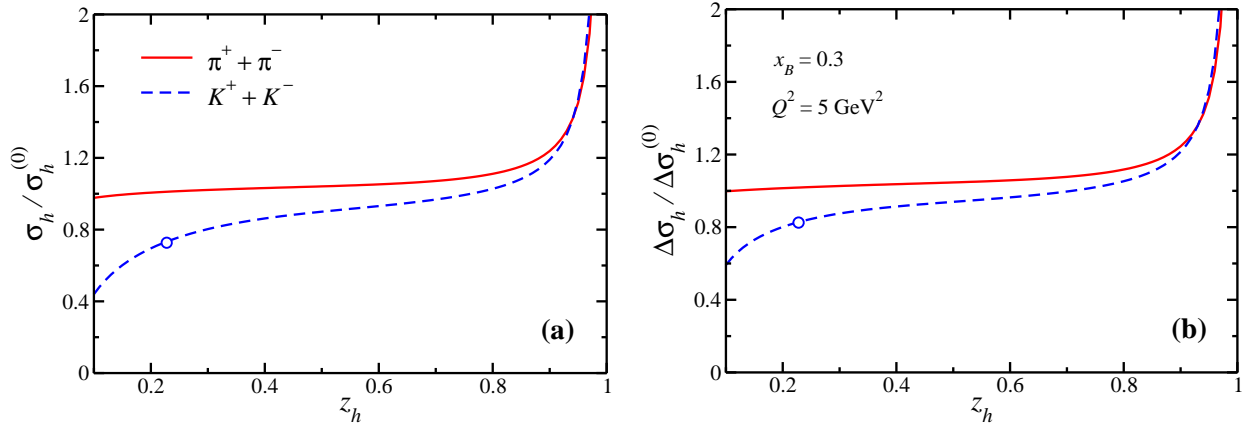

\includegraphics[width=8cm]{hmc_Hunp.eps}\ \ \
\includegraphics[width=8cm]{hmc_Hpol.eps}
\caption{Ratio of the (a) spin-averaged and (b) spin-dependent
	cross sections with and without HMCs for semi-inclusive
	production of various hadrons $h$ ($h = \pi^+ + \pi^-$
	or $K^+ + K^-$), for $x_B = 0.3$ and $Q^2 = 5$~GeV$^2$.
	The open circles denote the boundary between the target
	and current fragmentation regions.} 
\label{fig:had}
\end{figure}

While most of the existing SIDIS data have involved the production
of charged pions, the detection of heavier mesons and baryons can
provide complementary information on the flavor and spin structure
of PDFs, as well as on the dynamics of hadronization.
The production of kaons, for instance, tags strange or antistrange
quarks, and has been used with polarized targets as an independent
means of determining the $\Delta s$ distribution in the nucleon
\cite{HERMES_Deltas}, and for unpolarized scattering to determine
the magnitude of the $s$ distribution at small $x_B$ \cite{HERMES_s}.
In Fig.~\ref{fig:had} the spin-averaged and spin-dependent
cross section ratios with and without HMCs are shown for the
production of charged pions ($\pi^+ + \pi^-$) and kaons ($K^+ + K^-$)
at $x_B = 0.3$ and $Q^2 = 5~$GeV$^2$.
The effects are enhanced significantly with increasing hadron mass,
particularly at low $z_h$ values, mostly because of the
$(1+m_h^2/\zeta_h Q^2)$ factor in the $\xi_h$ variable in
Eq.~(\ref{eq:xi_h}).  Increasing values of $m_h^2/\zeta_h$ shift
the argument of the PDF to higher $x_B$, where the smaller magnitude
of the distributions effectively suppresses the cross section.

\begin{figure}[t]
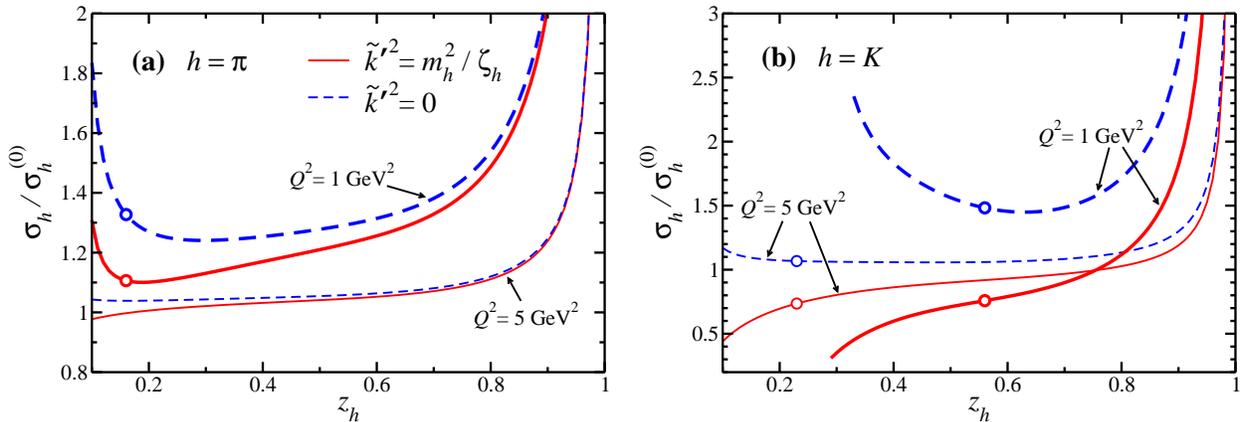

\vspace*{0.5cm}
\includegraphics[width=8cm]{hmc_kpr_pi.eps}\ \ \
\includegraphics[width=8cm]{hmc_kpr_K.eps}
\caption{Ratio of spin-averaged cross sections with and without HMCs
	for the production of (a) pions and (b) kaons, for different
	choices of the scattered parton invariant mass
	$\widetilde k'^2$ at $Q^2 = 1$~GeV$^2$ (thick lines)
	and $Q^2 = 5$~GeV$^2$ (thin lines) for $x_B = 0.3$. 
	The open circles denote the boundary between the target
	and current fragmentation regions.}
\label{fig:kpr}
\end{figure}

This phenomenon inherently arises from the choice of invariant mass
squared $\widetilde k'^2$ for the scattered quark, discussed in
Sec.~\ref{ssec:cf}.  While the choice of the mass becomes irrelevant
at high $Q^2$, the dependence on $\widetilde k'^2$ can be appreciable
at low $Q^2$ values, as Fig.~\ref{fig:kpr} illustrates.
Here the spin-averaged cross section ratios computed with
$\widetilde k'^2 = m_h^2/\zeta_h$ are compared with those
for massless partons, $\widetilde k'^2 = 0$, as used in
Ref.~\cite{Albino08}.
For the production of pions, the dependence on the quark virtuality
is negligible at $Q^2 = 5$~GeV$^2$, but becomes evident at lower
$Q^2$ for small $z_h$, $z_h \lesssim 0.5$.
Overall, the $\sigma_h/\sigma_h^{(0)}$ ratio is closer to unity
for the preferred choice of $\widetilde k'^2 = m_h^2/\zeta_h$
(see Sec.~\ref{ssec:cf}), with greater deviations
for the $\widetilde k'^2 = 0$ choice.

For kaon production the effects are expectedly larger, with
significant dependence on the quark virtuality at $Q^2 = 1$~GeV$^2$
for most of the $z_h$ range.  Interestingly, the sign of the
correction is different for the two choices of quark mass for
$z_h \lesssim 0.8$.  By $Q^2 = 5$~GeV$^2$ the dependence on
$\widetilde k'^2$ is weaker, except at $z_h \lesssim 0.3$ where
visible differences persist.  
These results suggest that care must be taken when extracting PDF
information from low-$Q^2$ SIDIS data at the extremeties of the $z_h$
spectra, particularly for heavier produced hadrons such as kaons.
Caution must also be exercised when including data in the target
fragmentation region at small $z_h$, where factorization based
on the use of fragmentation functions becomes more questionable.
In this region utilization of data may require the fracture
functions formalism as discussed in Refs.~\cite{Trentadue93,
Graudenz94, deFlorian95}.

\subsection{Mass corrections for specific experiments}

\begin{figure}[t]
\includegraphics[width=9cm]{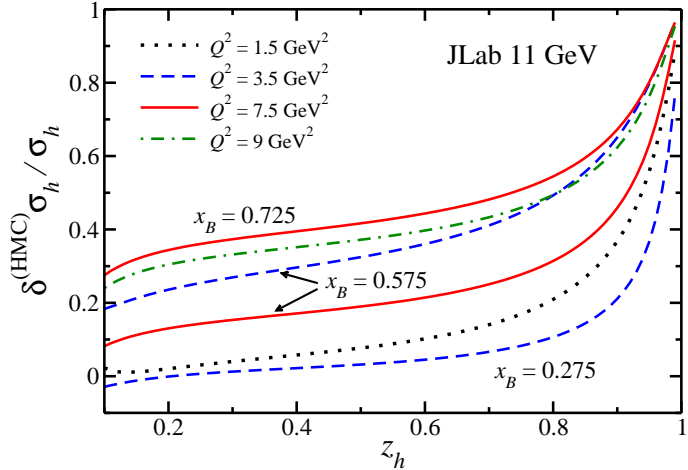}
\caption{Relative effect of HMCs on the spin-averaged SIDIS cross
	section	as a function of $z_h$ for $\pi$ production from
	protons	at kinematics typical of future 11~GeV Jefferson Lab
	experiments \cite{E12-09-007, E12-13-007, E12-06-109}.
	The relative HMC correction
	$\delta^{\rm (HMC)}\sigma_h / \sigma_h$
	is evaluated for each of the three $x_B$ values
	($x_B=0.275, 0.575$ and 0.725) at two fixed values of $Q^2$
	(indicated in the legend) from 1.5~GeV$^2$ to 9~GeV$^2$.}
\label{fig:exp_unp}
\end{figure}

\begin{figure}[t]
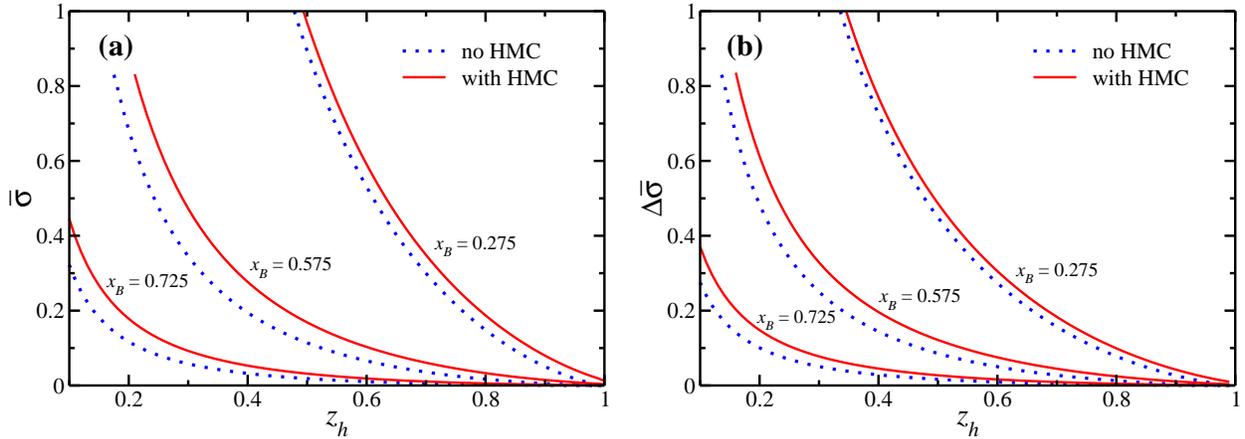

\vspace*{0.5cm}
\includegraphics[width=8cm]{hmc_unpCS.eps}\ \ \
\includegraphics[width=8cm]{hmc_polCS.eps}
\caption{Effect of HMCs on the
	(a) unpolarized and (b) polarized SIDIS cross sections
	for $\pi^+$ production from protons at typical 11~GeV
	Jefferson Lab kinematics \cite{E12-06-109}.
	The cross sections with (solid lines) and without
	(dotted lines) are evaluated at selected $x_B$
	(and corresponding $Q^2$) values,
	$x_B = 0.275$ ($Q^2 = 1.5$~GeV$^2$), 
	$x_B = 0.575$ ($Q^2 = 3.5$~GeV$^2$), and
	$x_B = 0.725$ ($Q^2 = 7.5$~GeV$^2$).}
\label{fig:exp_CS}
\end{figure}

The relevance of the HMCs to future SIDIS experiments is
illustrated in Fig.~\ref{fig:exp_unp}, where the difference
between the finite-$Q^2$ and Bjorken limit cross sections,
\begin{eqnarray}
\delta^{\rm (HMC)}\sigma_h
&=& \sigma_h - \sigma_h^{(0)},
\end{eqnarray}
is evaluated relative to the finite-$Q^2$ cross section at
kinematics typical of planned 11~GeV Jefferson Lab experiments
\cite{E12-09-007, E12-13-007, E12-06-109}.
The cross sections are computed using the same spin-averaged PDF
(CT \cite{CT10}) and fragmentation function (HKNS \cite{HKNS07})
parametrizations discussed in Sec.~\ref{ssec:kin} above.

The effects are pronounced mostly at large $z_h$, where
the ratio of uncorrected to corrected cross sections
$\sigma_h^{(0)} / \sigma_h \to 0$.  This is directly correlated
with the behavior of the fragmentation function ratio $D/D^{(0)}$
in Fig.~\ref{fig:FF}, which diverges as $z_h \to 1$ because at
finite $Q^2$ the fragmentation variable $\zeta_h < 1$ at $z_h=1$.
The effects at lower $z_h$ are stronger with increasing $x_B$ and
with decreasing $Q^2$.  In the range $0.3 \lesssim z_h \lesssim 0.6$,
which is typical for the coverage expected in the future experiments,
the HMCs are $\lesssim 10\%$ at $x_B = 0.275$ (where the corresponding
$Q^2$ is between 1.5 and 3.5~GeV$^2$), but increase to $\approx 40\%$
at $x_B = 0.725$ (for $Q^2$ between 7.5 and 9~GeV$^2$).

The effect of HMCs on the actual (reduced) spin-averaged
$\bar{\sigma}_h$ as well as spin-dependent $\Delta\bar{\sigma}_h$ 
cross sections is shown in Fig.~\ref{fig:exp_CS}, where the
cross sections are calculated at the same kinematics as in
Fig.~\ref{fig:exp_unp}.
The impact of HMCs on the cross sections are more pronounced for
increasing $x_B$ and decreasing $Q^2$.

\begin{figure}[t]
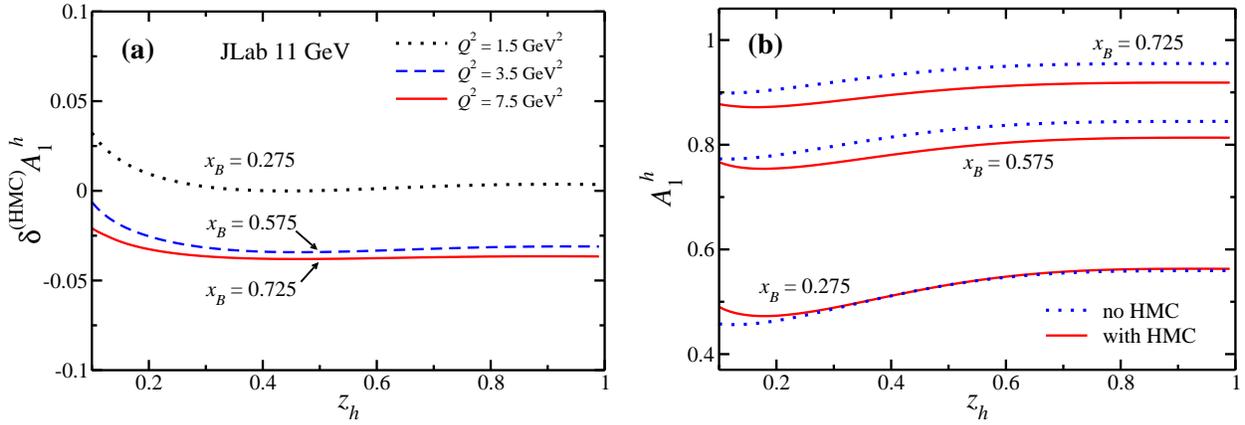

\vspace*{0.5cm}
\includegraphics[width=8cm]{hmc_expA1.eps}\ \ \
\includegraphics[width=8cm]{hmc_A1.eps}
\caption{Effect of HMCs on the SIDIS polarization asymmetry
	$A_1^h$ for $\pi^+$ production from protons, expressed as
	(a) relative shift $\delta^{\rm (HMC)} A_1^\pi$ and
	(b) effect on the asymmetry directly.
	The asymmetries are evaluated at typical 11~GeV
	Jefferson Lab kinematics \cite{E12-06-109}, for several
	values of $x_B$ and $Q^2$, as in Fig.~\ref{fig:exp_CS}.}
\label{fig:exp_pol}
\end{figure}

For spin-dependent scattering, the effects on the semi-inclusive
polarization asymmetry $A_1^h$ can also be quantified by defining
the difference with respect to the massless limit asymmetry
$A_1^{h (0)}$,
\begin{eqnarray}
\delta^{\rm (HMC)}A_1^h &=& A_1^h - A_1^{h (0)},
\end{eqnarray}
where (neglecting the transverse $g_2$ contribution)
	$A_1^h = A_\parallel^h/D$.
Here $A_\parallel^h = \Delta\sigma_h / 2\sigma_h$ is the parallel
asymmetry, and $D = (1 - (1-y) \varepsilon)/(1 + \varepsilon R)$
is the photon depolarization factor, with $R$ the ratio of
longitudinal to transverse photoproduction cross sections.
At leading order, the polarization asymmetry is then given in
terms of the ratios of sums of polarized and unpolarized PDFs,
\begin{eqnarray}
A_1^h &=&
\frac{\sqrt{1-\varepsilon^2}}{D}
\frac{\sum_q e_q^2\, \Delta q(\xi_h,Q^2)\, D_q^h(\zeta_h,Q^2)}
     {\sum_q e_q^2\, q(\xi_h,Q^2)\, D_q^h(\zeta_h,Q^2)},
\end{eqnarray}
where the kinematic prefactor
$\sqrt{1-\varepsilon^2}/D = \sqrt{1+\gamma^2}/(1+\gamma^2 y/2)$.
In the Bjorken limit this becomes unity, but at finite $Q^2$ it
represents the projection of the longitudinal lepton polarization
along the virtual photon direction,
$\cos\theta_{S_L} = D/\sqrt{1-\varepsilon^2}$, where $\theta_{S_L}$
is the angle between the lepton and photon spin vectors.

The corrections to the asymmetry $\delta^{\rm (HMC)}A_1^h$
as well as to the actual $A_1^h$ asymmetry are shown in
Fig.~\ref{fig:exp_pol} as a function of $z_h$ for the production
of $\pi^+$ mesons from a proton target, at fixed $x_B$ and $Q^2$
values corresponding to planned 11~GeV Jefferson Lab experiments
\cite{E12-06-109}.
At low $x_B$ values the differences are very small except at
very small $z_h$, where the effects increase as $z_h \to 0$.
As for the spin-averaged cross sections in Fig.~\ref{fig:exp_unp},
the effects increase with increasing $x_B$ and with decreasing $Q^2$.
At the highest $x_B$ value, $x_B = 0.725$, the asymmetry is reduced
by $\approx 0.03$ for $z_h \gtrsim 0.3$.  
If the asymmetry were to approach unity in the $x_B \to 1$ limit,
this would amount to a correction of $\approx 3\%-4\%$.
Somewhat larger corrections are obtained using the GJR \cite{GJR09}
and BB \cite{BB02} PDF parametrizations, although for this combination
the $A_1^h$ asymmetry is not guaranteed to respect the positivity
constraint at large $x_B$.  On the other hand, any dependence of the
HMCs on the input PDFs and fragmentation functions is in principle
artificial, since in the actual global analyses of SIDIS data the
distributions would be determined uniquely and self-consistently
by implementing the hadron mass corrected expressions in
Eqs.~(\ref{eq:dsig_both}) directly.

\section{Conclusion}
\label{sec:conclusion}

With the imminent completion of the 12~GeV upgrade of Jefferson Lab,
and with ongoing programs at existing facilities, a new generation of
SIDIS experiments will vastly improve our understanding of the spin
and flavor decomposition of parton distributions in the nucleon,
and explore the relatively new domain of transverse momentum
dependent parton distributions.
A full utilization of the new data will require that effects
associated with kinematical constraints at finite energy are
properly taken into account.
Following earlier work which studied the dependence of unpolarized
SIDIS cross sections on the masses of hadrons in the initial and
final states \cite{Mulders01, Albino07, AHM09}, in this work we have
presented a comprehensive analysis of hadron mass corrections to
both spin-averaged and spin-dependent cross sections and asymmetries
at finite values of $Q^2$.

Using the framework of collinear factorization, we have derived
formulas for SIDIS cross sections in the presence of HMCs, which
at leading order in $\alpha_s$ result in a rescaling of the PDFs
in terms of the modified Nachtmann variable $\xi_h$ and of the
fragmentation functions in terms of the finite-$Q^2$ fragmentation
variable $\zeta_h$.  Our results respect all kinematical limits
at finite $Q^2$, and reproduce the standard expressions in the
Bjorken limit.
An interesting feature of the modified formulas is that, in the
presence of HMCs, the parametric dependence on the scattering and
fragmentation variables in the parton distribution and fragmentation
functions becomes kinematically intertwined.  While this complicates
the analysis of SIDIS data in certain regions of kinematics,
the effects are calculable within our framework.

We have quantified the hadron mass effects numerically as a function
of the kinematic variables $x_B$, $z_h$ and $Q^2$ in order to
determine the regions where the corrections are most relevant.
Generally, the HMCs are strongest at large $x_B$ and low $Q^2$
(as for target mass corrections in inclusive DIS), and for large
as well as very low values of the fragmentation variable $z_h$.
The effects are also more dramatic for heavier hadrons such as
kaons than for pions at the same kinematics.
Extending the previous analysis of mass corrections in unpolarized
SIDIS \cite{AHM09}, we have also examined the dependence of the
HMCs on the choice of scattered parton virtuality, $\widetilde k'^2$.
In some cases the uncertainty due to this choice is quite
significant, particularly at small values of $Q^2$ and low $z_h$
for pions, and over a larger $z_h$ range for kaons, where the
correction is observed to change sign.

To illustrate the importance of HMCs in practical applications,
we have computed the corrections to SIDIS cross sections and
polarization asymmetries that would need to be applied at
kinematics relevant to upcoming experiments at Jefferson Lab
\cite{E12-09-007, E12-13-007, E12-06-109}.  Here the $x_B$
and $Q^2$ are necessarily correlated, so that usually the
data bins at small $x_B$ correspond to lower $Q^2$ values,
while at large $x_B$ the $Q^2$ is typically higher.
For unpolarized pion production, the HMCs are strongest at
large $z_h$, for all kinematics.  At intermediate $z_h$ values
the corrections at low $x_B$ are relatively small, $\lesssim 10\%$,
but increase to $\sim 40\%-50\%$ at higher $x_B$ ($x_B = 0.725$),
even at moderately large $Q^2$ ($Q^2 \approx 9$~GeV$^2$).
Qualitatively similar behavior is observed for the semi-inclusive
polarization asymmetry $A_1^h$, which receives larger HMCs at
higher $x_B$ values, although there is stronger sensitivity to
the specific behavior of the input PDFs.
Overall, our analysis suggests that mass corrections may be
an important ingredient in future analysis of SIDIS data from
facilities such as Jefferson Lab, especially at high values of
$x_B$, and particularly for hadrons heavier than the pion.

An immediate application of the results derived here will be in
upcoming global spin PDF analyses, such as by the JAM Collaboration
\cite{JAM}, which aims to fit an expanded set of high-energy
scattering data, including SIDIS, down to $Q^2 = 1$~GeV$^2$.
Future theoretical development of this work should include extending
the calculation to next-to-leading order in $\alpha_s$, which will
necessitate consideration of hadron production at nonzero transverse
momenta, $\bm{p}_{h \perp}$.  The work can also be extended to other
types of distributions measured in various single-spin asymmetries
in SIDIS reactions, which provide information not only on the
$x_B$ and $z_h$ distributions but also on the transverse momentum
of the partons.

\begin{acknowledgements}

We thank A.~Bacchetta, R.~Ent, T.~Hobbs and N.~Sato for helpful
discussions.
This work was supported by the DOE contract No. DE-AC05-06OR23177,
under which Jefferson Science Associates, LLC operates Jefferson Lab,
DOE contract DE-SC0008791, and NSF award No.~0653508.

\end{acknowledgements}
\newpage

\end{document}